\documentclass{amsart}
\usepackage{mathtools}
\usepackage[square,numbers]{natbib}
\usepackage{float}
\restylefloat{table}
\usepackage[utf8]{inputenc}
\usepackage[toc,page]{appendix}
\usepackage{amsmath,amsthm,amssymb,stmaryrd}
\usepackage{graphicx}
\usepackage[colorinlistoftodos,prependcaption]{todonotes}
\usepackage{xargs}
\usepackage{multicol}
\newcommandx{\unsure}[2][1=]{\todo[linecolor=blue,backgroundcolor=blue!25,bordercolor=blue,#1]{#2}}
\usepackage[section]{placeins}
\usepackage{float}
\usepackage{comment}
\usepackage{xcolor}
\usepackage{dcolumn}
\usepackage{bm}

\usepackage{graphicx} 
\usepackage{amsaddr}

\newcommand{\bbint}[2]{\ensuremath{\;\backslash\!\!\!\!\backslash\!\!\!\!\!\int_{#1}^{#2}}}

\title[Generalized Zeta Functions]{Regularization of Divergent Power Sums via Fractional Extension of Differential Generators}

\author{Eric A. Galapon}
\address{Theoretical Physics Group, National Institute of Physics, University of the Philippines, Diliman Quezon City, 1101 Philippines}
\date{\today}

\begin{document}
	
\maketitle
\begin{abstract} We reconsider the problem of regularizing  the divergent series $\sum_{n=1}^{\infty}n^{\alpha}$ for $\operatorname{Re}\alpha>-1$, and offer a regularization prescription that  yields the Riemann zeta regularization as a special case. The development of the regularization is framed as a two-step problem. The first step is prescribing a regularization of the divergent sum $\sum_{n=1}^{\infty}n^m$ for every non-negative integer $m$; and the second step is the extension of the sum for non-integer $\alpha$. The extension is obtained under the consistency condition that the regularized sum for integer $m$ emerges continuously from the sum for non-integer $\alpha$. The scheme is specified by a differential generator $L=L(\mathrm{d}/\mathrm{d}t)$ through which a generalized spectral function (GSF), $K_L(t)$, is constructed. Under the condition that the GSF has a holomorphic complex extension $K_L(z)$ with $z=0$ as a pole, the case for integer $m$ takes the regularized value $\sum_{n=1}^{\infty} n^m = (2\pi i)^{-1}\oint_C L^m K_L(z) z^{-1}\mathrm{d}z$, where $C$ is a closed contour enclosing only the pole of $K_L(z)$ at the origin. On the other hand, under the consistency condition, the case for non-integer $\alpha$ takes the value $\sum_{n=1}^{\infty}n^{\alpha}=(2\pi i)^{-1}\int_{\tilde{C}} L^{\alpha} K_L(z) z^{-1}\mathrm{d}z$, where $L^{\alpha}$ is the fractional extension of $L^m$ and $\tilde{C}$ is an appropriate deformation of the contour $C$. Here, we obtain the regularization corresponding to the generator $L=h(t) \mathrm{d}/\mathrm{d}t$, with $h(t)$ positive for all $t>0$, monotonically non-increasing, and admitting complex extension $h(z)$ such that $1/h(z)$ is entire. We find that the regularized sum is equal to the Riemann zeta regularized value plus terms determined by the generator $L$. 
\end{abstract}
	
\section{Introduction}
Many problems in physics require assigning a meaningful finite value to divergent infinite series of the form $\sum_{n=1}^{\infty}\lambda_n^{\alpha}$, where $\alpha$ is non-negative and $\lambda_n$ is a sequence of monotonically increasing positive numbers \cite{elizalde1,kirsten}. A prescription on how such a divergent series is assigned a value is referred to as a regularization. One such regularization is the zeta function regularization, which assigns value to the divergent series by analytic continuation \cite{elizalde1,kirsten}. In this prescription, the domain of the exponent $\alpha$ is extended to regions in the complex $\alpha$-plane where the series converges absolutely and is subsequently analytically continued from there into the entire complex plane. The divergent series is then assigned the value equal to the value of the analytic continuation at $\alpha$ at which the series diverges. The quintessential example and the namesake of the scheme is the regularization of divergent series $\sum_{n=1}^{\infty} n^{\alpha}$ for positive $\alpha$. Extending $\alpha$ to the region $\Re\alpha<-1$ lends the series convergent to the Riemann zeta function $\zeta(-\alpha)=\sum_{n=1}^{\infty}1/n^{-\alpha}$ which has a unique holomorphic extension in the entire complex plane. Since $\zeta(z)$ is analytic everywhere except at its simple pole $z=1$, $\zeta(-\alpha)$ is well defined for positive $\alpha$ and this value is assigned to the divergent series $\sum_{n=1}^{\infty} n^{\alpha}$. Hardy and Littlewood were the first to intimate that the Riemann zeta function can be used to assign value to the divergent sum $\sum_{n=1}^{\infty} n^{\alpha}$ \cite{hardy}. Later Hardy further developed the idea in his highly influential book ``Divergent Series'' and connected it to Ramanujan's summation method \cite{hardybook}. 

This viewpoint was later extended from ordinary series to the spectra of differential operators, where one considers zeta functions formed from the eigenvalues of an operator. Such spectral zeta functions and their analytic continuation were first developed in the work of Minakshisundaram and Pleijel \cite{minakshisundaram}, and were subsequently placed on a rigorous analytic footing by Seeley via his theory of complex powers of elliptic operators \cite{seeley}. In mathematics, the interpretation of the determinants of differential operators through spectral zeta functions was first developed systematically by Ray and Singer in their work on analytic torsion \cite{ray}. The first explicit application of the zeta‑function regularization to a physical problem is due to Dowker and Critchley in 1976 where they use the analytic continuation of the generalized zeta‐function of an elliptic operator to regulate one‐loop determinants and stress tensors in curved spacetime \cite{dowker}. Shortly thereafter, in 1977, Hawking independently adopted the same idea for path–integral quantization, in particular, the evaluation of partition function on curved backgrounds \cite{hawking}. Zeta-function regularization has continued to evolve, with ongoing refinements and extensions in both mathematical formulation \cite{candelpergher,tao} and physical applications \cite{padilla}.   

Although zeta-function regularization has proven highly effective in these contexts, its conventional formulation leaves open the possibility that certain physically relevant effects may lie beyond its scope. A viable example arises in the case of non-interacting fermions in a box at zero temperature, where additional regularization beyond the conventional framework may be necessary. Suppose a box is filled with non-interacting fermions in the thermodynamic limit of infinitely many fermions. The possible energies of a particle of mass $m$ in a box  of length $L$ are $E_n=n^2 h^2/8 m L^2$ for every non-negative integer $n$.  Then, by Pauli's exclusion principle, a pair of fermion occupies every energy state, so that the total energy is $(h^2/4mL^2)\sum_{n=1}^{\infty}n^2$. It is divergent and if we regularize using zeta function regularization the result is $(h^2/4mL^2)\zeta(-2)$, which vanishes because $\zeta(-2)=0$. 

The vanishing of the regularized energy may not be an issue at all on its own. But now suppose we insert a non-porous partition that divides the box in two compartments each of length $L/2$, keeping both sides filled with non-interacting fermions. As long as the widths of the two partitions are equal, the net force on the partition is zero. Suppose now that the partition is displaced by some $x$. It is not unreasonable to expect that an energy density imbalance is created between the two sides and a restoring force arises, which is calculated to be 
\begin{equation}
    F=-\frac{48 h^2}{m L^4}\left(x+O(x^3)\right)\sum_{n=1}^{\infty}n^2 .
\end{equation}
The Riemann zeta regularization of the divergent series leads to no restoring force because $\zeta(-2)=0$. It is not clear at all why and how the restoring force would vanish. However, it is more coherent to suspect that the restoring force is not equal to zero. This apparent inconsistency indicates that the regularized sum alone may not capture all physically relevant features of the system. Earlier studies on the Casimir effect in analogous systems show that standard regularization may fail to take into account physically relevant features in certain regimes \cite{volovik,kolomeisky}, thereby motivating the exploration of alternative regularization schemes that may encode information beyond the spectrum.

In this paper, we develop such a scheme---a regularization that supplements the standard zeta-regularization with spectrum independent data and recovers Riemann regularization as a special case. We will limit ourselves to the extension of the zeta-regularization of the divergent series $\sum_{n=1}^{\infty}n^{\alpha}$. Our method proceeds from the observation that the zeta function regularization of the sum $\sum_{n=1}^{\infty} n^m$ for every non-negative integer $m$, known as the trace identities,  can be obtained from the spectral function $K(t)=\sum_{n=1}^{\infty} e^{-n t}$ for the Riemann zeta function. Clearly $(-D_{t})^m K(\tau)=\sum_{n=1}^{\infty} n^m e^{-n t}$, where $D_{t}$ is differentiation with respect to $t$. The sum converges for all $t>0$ but diverges as $t\rightarrow 0$. But in the limit, we formally obtain  the divergent sum $\sum_{n=1}^{\infty}n^m$ which are the divergent values of the trace identities. A regularization of the divergent sum is obtained by an appropriate identification of the terms that diverge in the limit as $t\rightarrow 0$. This is done by explicitly summing the spectral function, giving $K(t)=(e^{t}-1)^{-1}$, 
followed by expanding it in powers of $t$, yielding $K(t)=\sum_{k=0}^{\infty}\frac{B_k}{k!} t^{k-1}$ where the $B_k$'s are the Bernoulli numbers. Then acting $(-D_{t})^m$ on the expansion yields the Laurent series expansion of $(-D_{\tau})^m K(t)$ about $t=0$.   The desired regularization is obtained by taking the limit of the regular part of the expansion as $t\rightarrow 0$ and assigning the limit as the value of the divergent series $\sum_{n=1}^{\infty}n^m$. This returns the value $(-1)^m B_{m+1}/(m+1)$, which is precisely equal to $\zeta(-m)$. Then we obtain the regularized trace identities
$\sum_{n=1}^{\infty} n^m \equiv \zeta(-m)$,
the value assigned by the Riemann zeta function regularization. 

The crucial observation is that the spectral function $K(t)$ is built from the solutions to $-D_{t}g_n(t)=n g_n(t)$ subject to the boundary condition $g_n(0)=1$ for every positive integer $n$, which are $g_n(t)=e^{-n t}$, from which $K(t)=\sum_{n=1}^{\infty} g_n(t)$. This offers an avenue in generalizing the Riemann zeta regularization of trace identities by introducing a generator $L$ of a regulator, which we will denote by $\mathcal{R}_L(m)$. For the Riemann zeta regularization, the generator is $L_0=-D_{t}$ and the regulator is $\mathcal{R}_{L_0}(m)=\zeta(-m)$. In this paper, we develop this idea of constructing regulators from a generator. We will frame the problem as a two-step problem. The first step is prescribing a regularization of the divergent sum $\sum_{n=1}^{\infty}n^m$ for all non-negative integers $m$, which we will refer to as the integral trace identities, corresponding to a given generator $L$; and the second step is the extension of the sum to non-integer values of $m=\alpha$, which we will refer to as the fractional trace identity. Once the case of the integral trace identities have been solved, we solve the case of the fractional trace identity non-integer powers by means of extending the positive integer powers of $L$, $L^m$, into fractional powers of $L$, $L^{\alpha}$. We require that the case of integer powers arises continuously as a limit of the fractional-power case. 

In this paper, we will investigate in detail the regularization arising from the family of generators
\begin{equation}\label{generator}
    L=-h(t)D_t ,
\end{equation}
with $h(t)$ positive for all $t>0$, monotonically non-increasing, and admitting complex extension $h(z)$ such that $1/h(z)$ is entire. The well-known Riemann zeta function regularization corresponds to $h(t)=1$. For other choices, the regulator takes the form $\mathcal{R}_L(\alpha)=\zeta(-\alpha)+correction$ for integer or non-integer $\alpha$, where the correction term is a finite-part of a divergent integral determined by $h(t)$.

The remainder of the paper is organized as follows. In Section-\ref{framework}, we formulate a generalization of the Riemann zeta function regularization based on a prescribed generator $L$. In Section-\ref{gentrace}, we obtain the regularization of the integral trace identities. In Section~\ref{fractionaltrace}, we extend the construction to fractional trace identities by introducing the fractional operator $L^{\alpha}$, which we obtain in both differential and integral representations. In Section~\ref{hankelzeta}, we construct regulators defined by contour integrals along the Hankel contour. Finally, in Section \ref{conclusion} we discuss prospective developments, including the obstruction to extending the present results, the path toward a rigorous formulation, and the insights that may emerge from the proposed generalized zeta regularization.

\section{The Framework}\label{framework}
To a particular regularization, we associate a generator $L$ that admits a solution to the eigenvalue problem 
\begin{equation}
    L g_n(t)=n g_n(t)
\end{equation}
for every positive integer $n$. In parallel to the development of the zeta spectral function, we define the generalized spectral function (GSF) corresponding to $L$ as
\begin{equation}\label{gsforg}
    K_L(t)=\sum_{n=1}^{\infty} g_n(t) .
\end{equation}
We demand that $K_L(t)$ diverges at $t=0$ and that the infinite series \eqref{gsforg} converges uniformly for all $t>0$ so that $L^m K_L(t)=\sum_{n=1}^{\infty} n^m g_n(t)$; moreover, it must satisfy
\begin{equation}\label{quack}
\left.L^m K_L (t)\right|_{t=0}=\sum_{n=1}^{\infty} n^m.
\end{equation}
These impose the boundary conditions on the eigenfunctions of $L$,
\begin{equation}
    g_n(0)=1,\;\;\; g_n(\infty)=0 .
\end{equation}

It is the property \eqref{quack} that allows $K_L(t)$ to define a regulator for the divergent series $\sum_{n=1}^{\infty}n^m$.  
A regularization, to be denoted by $\mathcal{R}_L(m)$, is obtained by expanding $K_L(t)$ about $t=0$, keeping only the constant term of the expansion to assign as the regularized value of the trace identity. This is typically taken as the finite-part (fp) of the expansion to yield the regularization of the trace identities \cite{kirsten}, 
\begin{equation}\label{fpiinteg}
    \mathcal{R}_L(m)=\mathrm{fp}\left[L^m K_L(t)\right]_{t=0} .
\end{equation}
It will become apparent later that a different interpretation other than the finite-part is necessary. As we have demanded above, we require that the regularization must be continuous. That is the regularization for $\sum_{n=1}^{\infty}n^{\alpha}$ for non-integer $\alpha$ must continuously approach the integral trace identities $\sum_{n=1}^{\infty}n^m$ as $\alpha$ approaches non-negative integer values $m$. Keeping the interpretation as a finite-part will generally not satisfy this condition. We shall establish that the way to satisfy the continuity condition is to start from the integral trace identities and from them work our way to obtain the regularization in the non-integral case. 

The key requirement to accomplish this is to impose that $K_L(t)$ inherits the behavior of the spectral zeta function $K(t)$ at $t=0$: possessing a holomorphic extension throughout the complex $t$-plane, with a simple pole at $t=0$. This allows us to extract the constant term in $L^m K_L(t)$ using the Cauchy integral formula, in particular, by means of the contour integral 
\begin{equation}\label{contourinteg}
    \mathcal{R}_L(m) = \frac{1}{2\pi i}\oint_C L^m K_L(z)\,\frac{\mathrm{d}z}{z},
\end{equation}
where the contour $C$ encloses the pole $z=0$ without enclosing any other singularities of $L^m K_L(z)$. In the language of finite-part integration \cite{galapon2017PRSA,ticagalapon2018JMP,ticagalapon2019JMP,villanuevagalapon,galapon2023AA,ticagalapon2023PRSA,ticablancasgalapon2024PRSA,blancasgalapon2025}, the regularization \eqref{contourinteg} is the regularized limit of $L^m K_L(t)$ as $t\rightarrow 0$ \cite{galapon2023AA}. If we only wish to regularize the trace identities, then the prescriptions \eqref{fpiinteg} and \eqref{contourinteg} are equivalent when $z=0$ is an isolated singularity of $K_L(t)$. However, under the condition that the integral case emerges from the fractional case continuously.  

The extension to non-integer powers is realized by replacing the integer powers $L^m$ with a suitable fractional power $L^{\alpha}$. This substitution alters the analytic behavior of the integrand, i.e. the isolated poles of $L^m K_L(z)$ are replaced, in general, by branch points in $L^{\alpha} K_L(z)$. The corresponding regularization is then obtained by deforming the contour of integration to arrive at the representation 
\begin{equation}
    \mathcal{R}_L(\alpha) = \frac{1}{2\pi i} \int_C L^{\alpha} K_L(z) \frac{\mathrm{d}z}{z}
\end{equation}
where the contour is chosen to avoid intersecting any branch cuts and to ensure analyticity of the integrand along the path. Moreover, the contour is required to depend continuously on $\alpha$ in such a way that, in the limit $\alpha\rightarrow m \in \mathbb{N}$, the branch cuts collapse and the contour reduces to a closed loop encircling the origin, thereby recovering the integer case.

\section{Generalized Zeta Function Regularization of the Integral Trace Identities}\label{gentrace}
We now develop a generalization of the Riemann zeta regularization of the divergent trace identities  $\sum_{n=1}^{\infty} n^{m}$ arising from the generator $L=-h(t) D_t$ in \eqref{generator}. The generalized spectral function is obtained by solving the eigenvalue problem,
\begin{equation}
		-h(t) g_n'(t)= n g_n(t) ,
\end{equation}
for every positive integer $n$, subject to the conditions $g_n(0)=1$ and $g_n(\infty)=0$. The solution is given by
\begin{equation}\label{solution}
		g_n(t)=e^{-n \Phi(t)} .
\end{equation}
where
\begin{equation}\label{myphi}
    \Phi(t)=\int_0^t \frac{\mathrm{d}u}{h(u)}.
\end{equation}
Under the conditions on $h(t)$, $\Phi(t)$ exists for all $t\geq 0$; moreover, it is positive due to the positivity of $h(t)$. Substituting the solution \eqref{solution} back into equation \eqref{gsf} and evaluating the sum, we obtain the corresponding generalized spectral function
\begin{equation}\label{gsf}
		K_L(t)=\frac{1}{e^{\Phi(t)}-1} .
\end{equation}
Comparing the generalized spectral function \eqref{gsf} with the zeta function spectral function $K_{L_0}(t)=(e^t-1)^{-1}$, we observe that the generator induces the transformation $t\mapsto\Phi(t)$ at the level of the spectral function $K_{L_0}(t)$. Since $\Phi(0)=0$ and $\Phi'(0)\neq 0$, the origin $t=0$ is a simple zero of $\Phi(t)$ so that $K_L(t)$ has a simple pole there. Hence, the transformed spectral function $K_L(t)$ retains the same holomorphic behavior as $K_{L_0}(t)$ at $t=0$. This will play an important role in the development to follow.
	
To obtain the $m$-th action of the generator $L$, we perform a change in variable from $t$ to $\tau=\exp(-\Phi(t))$, with $\tau$ taking values in the range $(0,1)$. This leads to the equality
\begin{equation}
		\left(-h(t)D_t\right)^m\frac{1}{e^{\Phi(t)}-1} = \left(\tau D_{\tau}\right)^m \frac{\tau}{1-\tau} .
\end{equation}
We compare this with the known representation of the polylogarithm function, $\operatorname{Li}_{\nu}(z)$, with order  $\nu=-m$, for positive integer $m$,
\begin{equation}
		\operatorname{Li}_{-m}(z)=\left(z D_z\right)^m \frac{z}{1-z} .
\end{equation}
The polylogarithm $\operatorname{Li}_{-m}(z)$ has a pole of order $(m+1)$ at $z=1$ and it is explicitly given by
\begin{equation}\label{polym}
		\operatorname{Li}_{-m}(z)=\frac{1}{(1-z)^{m+1}} \sum_{k=0}^{m-1}\left<{m\atop k}\right> z^{n-k},
\end{equation}
where $\left<{m\atop k}\right>$ are the Eulerian numbers. Thus we have established the following result,
	\begin{equation}\label{integercase}
		\left(-h(t)D_t\right)^m K_L(t)  = \operatorname{Li}_{-m}\left(e^{-\Phi(t)}\right) .
	\end{equation}
for all positive integer $m$. 
	
Our regularization requires analytically extending $\operatorname{Li}_{-m}(e^{-\Phi(t)})$ to the complex plane to implement the contour integration in equation \eqref{contourinteg}. This demands that $\Phi(t)$ has a complex extension $\Phi(z)$ that is analytic in the neighborhood of $z=0$. This is ensured by the restriction that $h(t)$ has a complex extension $h(z)$ such that $1/h(z)$ is entire. Then, for complex $z$, $\Phi(z)$ is given by equation \eqref{myphi} with $t$ replaced by $z$ and the contour of integration is any simple path connecting the origin and the point $z$. Since the integral exists for all $z$ and $1/h(z)$ is entire, $\Phi(z)$ is itself entire. This implies that $e^{-\Phi(z)}$ is also entire. Then, from equation \eqref{polym}, $\operatorname{Li}_{-m}(e^{-\Phi(z)})$ has a pole at $z=0$ of order $m+1$ and analytic everywhere in the finite complex plane. The constant term is then obtained by performing a closed contour integration around the origin, the value of which is the desired regulator
\begin{equation}\label{rit}
\mathcal{R}_L(m)=\frac{1}{2\pi i} \oint_C \operatorname{Li}_{-m}\left(e^{-\Phi(z)} \right)\frac{\mathrm{d}z}{z} .
\end{equation}

We unravel equation \eqref{rit} and establish the relation between Riemann zeta regularization and generalized zeta function regularization. To do so, we use the identity
\begin{equation}\label{polylogexp}
    \operatorname{Li}_{s}\left(e^{\mu}\right)= \Gamma(1-s) (-\mu)^{s-1} + \zeta(s) + \sum_{k=1}^{\infty} \frac{\zeta(s-k)}{k!} \mu^k 
\end{equation}
valid for $|\mu|<2\pi$ and $s\neq 1, 2, 3,\dots$ \cite{polylogsumrep}. We set $\mu=-\Phi(z)$. Since $\Phi(0)=0$, the continuity of $\Phi(z)$ at $z=0$ ensures that, for $z$ sufficiently close to zero, $|\mu|=|\Phi(z)|$ can be made arbitrarily small. This guaranties that the expansion remains valid in a neighborhood of $z=0$. Substituting $-\Phi(z)$ for $\mu$ yields
\begin{equation}\label{pwekpwek}
    \operatorname{Li}_{-m}\left(e^{-\Phi(z)}\right)= m! \Phi(z)^{-m-1} + \zeta(-m)+\sum_{k=1}^{\infty}(-1)^k \frac{\zeta(-m-k)}{k!} \Phi(z)^k .
\end{equation}

Equation \eqref{pwekpwek} is then substituted back into equation \eqref{rit}. By construction $\Phi(z)$ has a simple zero at the origin, so $\Phi(z)^k$ has zero of order $k$ at the origin. Then $\Phi(z)^k$ has no (non-zero) constant term for all positive integers $k$ so that each term in the infinite series does not contribute in the contour integral. On the other hand, the first term $m! \Phi(z)^{-m-1}$ may have a constant term for a given non-negative integer $m$ depending on the underlying $h(t)$, so that it may contribute to the integral. Finally, the second term is a constant term, so it definitely contributes. Then the regulator assumes the form 
\begin{equation}\label{xo}
\mathcal{R}_L(m) = \zeta(-m) + \frac{m!}{2\pi i}\oint_{C} \frac{1}{\Phi(z)^{m+1}}\frac{\mathrm{d}z}{z} .
\end{equation}
The contour integral in \eqref{xo} can be rewritten following the observation that $\Phi(z)$ has a simple zero at $z=0$ which implies that $z/\Phi(z)$ is analytic at $z=0$. This leads to equality
\begin{equation}
    \frac{m!}{2\pi i}\oint_{C} \frac{1}{\Phi(z)^{m+1}}\frac{\mathrm{d}z}{z}= \frac{1}{m+1} \left.\frac{\mathrm{d}^{m+1}}{\mathrm{d}z^{m+1}}\left(\frac{z}{\Phi(z)}\right)^{m+1}\right|_{z=0},
\end{equation}
which derives from the Cauchy integral formula for the derivative. Then the regulator becomes 
\begin{equation}\label{intheorem}
\mathcal{R}_L(m) = \zeta(-m) + \frac{1}{m+1} \left.\frac{\mathrm{d}^{m+1}}{\mathrm{d}z^{m+1}}\left(\frac{z}{\Phi(z)}\right)^{m+1}\right|_{z=0},
\end{equation}
for every non-negative integer $m$ .

When $h(t)=1$, we have $\Phi(z)=z$ so that $z/\Phi(z)=1$, which has zero derivative for all $m$ so that the contour integral in equation \eqref{intheorem} vanishes. Then the regularization reduces to the Riemann zeta regularization of the trace identities. However, for $h(t)\neq 1$, the contour integral generally does not vanish for all $m$, and it can be evaluated explicitly to give the following first trace identities.
\begin{eqnarray}
	&&\sum_{n=1}^{\infty}1 = \zeta(0)+\frac{1}{2} h'(0)\nonumber, \\
	&&\sum_{n=1}^{\infty} n = \zeta(-1) + \frac{1}{12} \left(4 h(0) h''(0)+h'(0)^2\right)\nonumber,\\
	&&	\sum_{n=1}^{\infty}n^2 = \zeta(-2)+\frac{1}{4} \left(h^{(3)}(0) h(0)^2+2 h(0) h'(0) h''(0)\right)\nonumber,\\
	&&	\sum_{n=1}^{\infty} n^3=\zeta(-3)+\frac{1}{120} \left(24 h(0)^3 h^{(4)}(0)+56 h(0)^2
	h''(0)^2\right.\nonumber\\
	&&\hspace{24mm}\left.-h'(0)^4+108 h(0)^2 h^{(3)}(0) h'(0)+64 h(0) h'(0)^2
	h''(0)\right)\nonumber
\end{eqnarray}
Alternatively, these can be obtained by repeated application of $L$ $m$-times on $K_L$, followed by evaluating the regularized limit as $z\rightarrow 0$. 

Observe that the regularized trace identities depend only on the values of the derivatives of $h(t)$ at $t=0$. For a given $m$, the regularized value of $\sum_{n=1}^{\infty} n^m$  depends only on the first $m$-derivatives of $h(t)$ at the origin. Thus, different regularizations corresponding to different $h(t)$'s will assign the same regularized value to the divergent sum if the first $m$ derivatives of the $h(t)$'s coincide at the origin.

\section{Regularization of the Fractional Trace Identity}\label{fractionaltrace}
We now wish to extend our result to the fractional trace identity $\sum_{n=1}^{\infty}n^{\alpha}$ for non-integer $\alpha$ with $\operatorname{Re}\alpha>-1$. Equation \eqref{integercase} offers the possibility of extending the integral trace identities to non-integer $m$ by mere replacement of $m$ by non-integer value $\alpha$. But such substitution without justification is ad hoc and hardly insightful. We proceed by requiring that the extension be derived from an appropriate fractional extension of the differential operator $L^m=(-h(t)D_t)^m$. We demand that such fractional extension satisfies the condition
\begin{equation}\label{consistency}
		\left(-h(t)D_t\right)^{\alpha} g_n(t) = n^{\alpha} g_n(t) ,
\end{equation}
for arbitrary complex $\alpha$ and for all $t>0$. This reduces to the case of integer powers of $L$ when $\alpha$ takes integer values. It is known that there is no unique fractional extension of an operator, but we will show that condition \eqref{consistency} is sufficient to fix the fractional extension.

\subsection{The Fractional Extension}
 We will obtain the fractional extension in two ways, in differential and integral forms. These two are completely different, but we will see that they will yield the same results. The two methods offer two different avenues to extend the results presented in this paper.

\subsubsection{In differential form}
We proceed with the fact that we have established the following identity
\begin{equation}
	\left(-h(t) D_t\right)=z D_z
\end{equation}
under the transformation performed above. The following identity is known
\begin{equation}
		(z D_z)^m = \sum_{k=0}^m \left\{{m\atop k}\right\} z^k D_z^k , \;\; m=0, 1, 2, \dots ,
\end{equation}
where $\left\{{m\atop k}\right\}$ are the Stirling numbers of the second kind \cite{knopp}. From this we define the complex differentiation of order $\alpha$ by replacing $m$ with $\alpha$ and extending the limit to infinity,
\begin{equation}\label{fractionalextension}
	(z D_z)^{\alpha}=\sum_{k=0}^{\infty}\left\{{\alpha\atop k}\right\} z^k D_z^k , \;\; \alpha\in\mathbb{C} .
\end{equation}
	
This requires extending the Stirling number to complex arguments. There is no single way to do this. At least two different definitions have been proposed \cite{richmond,flajolet}. We chose by means of the contour integral representation of the Stirling numbers of the second kind,
\begin{equation}
		\left\{m \atop k\right\} =\frac{m!}{k!} \frac{1}{2\pi i}\oint_C \left(e^{z}-1\right)^k \frac{\mathrm{d}z}{z^{m+1}},
\end{equation}
where $C$ is a contour that surrounds the origin, which is precisely the extension given in \cite{flajolet}. For complex $m=\alpha$, the origin becomes a branch point and we take the principal value of $z^{-\alpha-1}$. The contour is then taken to be the Hankel contour starting from $-\infty$ below the real line, going around the origin to go back to $-\infty$ from the top of the line. This leads to the extension
\begin{equation}
	\left\{\alpha \atop k\right\} =\frac{1}{k!} \sum_{l=1}^{k} (-1)^{k-l} \binom{k}{l} l^{\alpha},\;\; k\in \mathbb{N}^+,\; \alpha\in\mathbb{C} .
\end{equation}
	
This choice is justified if the fractional operator \eqref{fractionalextension} satisfies equation \eqref{consistency} or equivalently
\begin{equation}\label{xixi}
	\left(z D_z\right)^{\alpha} z^n = n^{\alpha} z^n ,
\end{equation}
for all positive integers $n$. We now show that this is indeed the case. Evaluating $(z D_z)^{\alpha}z^n$, we obtain after some simplifications the following
\begin{equation}
		(z D_z)^{\alpha} z^n =z^n \sum_{k=1}^{n}\sum_{l=1}^k (-1)^{k-l}\binom{k}{l}\binom{n}{k} l^{\alpha}  .
\end{equation}
To evaluate the double series, we gather terms that correspond to the same index $l$. This yields the equality
\begin{equation}
	\sum_{k=1}^{n}\sum_{l=1}^k (-1)^{k-l}\binom{k}{l}\binom{n}{k} l^{\alpha}=\sum_{l=1}^n \sum_{k=l}^n (-1)^{k-l}\binom{k}{l} \binom{n}{k} l^{\alpha}.
	\end{equation}
The coefficient of $l^{\alpha}$ for a fixed $l$ can be further simplified by shifting the index $k$ to $k-l$ and expanding the binomial coefficients, 
\begin{equation}
\sum_{k=l}^n (-1)^{k-l}\binom{k}{l} \binom{n}{k}=\frac{n!}{l!} \sum_{j=0}^{n-l}\frac{(-1)^j}{(n-j-l)! j!}
\end{equation}
The coefficient of $n^{\alpha}$, that is for $l=n$, evaluate to 1, while the rest of the coefficients vanish. Thus we have established the desired property \eqref{xixi}.

We are now ready to obtain the fractional version of equation \eqref{integercase}. We use the transformed variable $\tau$. Recall that $\tau$ fall in the interval $(0,1)$. In this variable,
\begin{equation}
	(-h(t)D_t)^{\alpha}K_L(t)=\left(\tau D_{\tau}\right)^{\alpha}\frac{\tau}{1-\tau}.
\end{equation}
For all $\tau$ we can expand $\tau(1-\tau)^{-1}=\sum_{k=1}^{\infty}\tau^k$. We substitute this back and using the result \eqref{xixi} we arrive at
\begin{equation}
		\left(\tau D_{\tau}\right)^{\alpha}\frac{\tau}{1-\tau}=\sum_{k=1}^{\infty} k^{\alpha} \tau^k .
\end{equation}
The infinite series converges for all values of $\tau$ in $(0,1)$ and complex $\alpha$, and evaluates to the polylogarithmic function $\operatorname{Li}_{-\alpha}(\tau)$. Going back to the original variable $t$, we establish
\begin{equation}
\left(-h(t) D_t\right)^{\alpha} K_L(t)=\operatorname{Li}_{-\alpha}\left(e^{-\Phi(t)}\right) .
\end{equation}
This is just the result of replacing the integer $m$ with the complex number $\alpha$.
	
\subsubsection{In integral form} We now extend the differential operator to fractional order in integral form \cite{miller}. Given some function $f(t)$, we have the action
	$-h(t) D_t f(t)=g(t)$.
We formally obtain the inverse
\begin{equation}
\left(-h(t)D_t\right)^{-1}=-\int_a^t \frac{\mathrm{d}y}{h(y)}
\end{equation}
for some $a$ to be specified later. We demand that this is both a left and right inverse. Indeed
\begin{equation}
	(-h(t) D_t)(-h(t)D_t)^{-1}f(t)=f(t),
\end{equation}
but
\begin{equation}
(-h(t)D_t)^{-1}(-h(t) D_t)f(t)=f(t)-f(a),
\end{equation}
We then restrict the domain of $-h(t)D_t$ to those satisfying $f(a)=0$. By induction
\begin{equation}
(-h(t)D_t)^{-n} f(t)=\frac{(-1)^n}{(n-1)!}\int_a^t \frac{f(y)}{h(y)}\left(\int_y^t\frac{\mathrm{d}s}{h(s)}\right)^{n-1}\,\mathrm{d}y
\end{equation}
We chose $a$ so as to eliminate the factor $(-1)^n$ to avoid introducing complex numbers at this point. This is accomplished with the choice $a=\infty$. The boundary condition on the domain is now $f(\infty)=0$; this is the origin of the boundary condition $g_n(\infty)=0$ on the eigenfunctions of $L$. Then
\begin{equation}
	(-h(t)D_t)^{-n} f(t)=\frac{1}{\Gamma(n)}\int_t^{\infty} \frac{f(y)}{h(y)}\left(\int_t^y\frac{\mathrm{d}s}{h(s)}\right)^{n-1}\,\mathrm{d}y .
\end{equation}
The desired fractional extension is obtained with the replacement $n\rightarrow -\alpha$,
\begin{equation}
(-h(t)D_t)^{\alpha} f(t)=\frac{1}{\Gamma(-\alpha)}\int_t^{\infty} \frac{f(y)}{h(y)}\left(\int_t^y\frac{\mathrm{d}s}{h(s)}\right)^{-\alpha-1}\,\mathrm{d}y .
\end{equation}
The outer integral converges provided $\alpha<0$. The region does not encompass the entire regularization region which is $\alpha>-1$. It can only deal in the range $(-1,0)$. This is in sharp contrast with the differential which holds in the entire regularization region. This is remedied by analytic continuation which we will do shortly.
	
In the mean time, it is sufficient for us to work in the region $(-1,0)$. We demand that the extension satisfy the condition \eqref{consistency}, in particular,
\begin{equation}\label{concon}
	(-h(t)D_t)^{\alpha} e^{-n \Phi(t)} = n^{\alpha} e^{-n \Phi(t)} ,\;\; \alpha<0 ,
\end{equation}
for all positive integer $n$. We have
\begin{equation}\label{pe}
(-h(t)D_t)^{\alpha}g_n(t) = \frac{1}{\Gamma(-\alpha)} \int_t^{\infty} e^{-n \Phi(y)} \left(\int_t^y\frac{\mathrm{d}s}{h(s)}\right)^{-\alpha-1}\,\frac{\mathrm{d}y}{h(y)} .
\end{equation}
This is evaluated by changing variable from $y$ to $s=\int_0^y\mathrm{d}t/h(t)$, followed by another change in variable from $s$ to $u=s-s(x)$. The right hand side of \eqref{pe} becomes
\begin{equation}
	\frac{1}{\Gamma(-\alpha)}e^{-ns(x)}n^{\alpha}\int_0^{\infty} e^{-\tau} \tau^{-\alpha-1}\,\mathrm{d}\tau .
\end{equation}
Under the condition that $\alpha$ is in the interval $(-1,0)$, the integral evaluates to $\Gamma(-\alpha)$. Substituting this back into equation \eqref{pe} we verify the equality \eqref{concon}.
	
Let us now look into the action of the extension to the generalized spectral function. Acting the fractional extension,
\begin{equation}
(-h(t)D_t)^{\alpha} K_L(t)=\frac{1}{\Gamma(-\alpha)} \int_t^{\infty} \frac{\mathrm{d}y}{h(y)}\left(\int_t^y \frac{\mathrm{d}x}{h(x)}\right)^{-\alpha-1} \left({e^{\Phi(y)}}-1\right)^{-1} .
\end{equation}
Again to evaluate this, we perform the same series of variable change performed above. This yields
\begin{equation}\label{pek}
\frac{e^{-s(t)}}{\Gamma(-\alpha)} \int_0^{\infty} \frac{u^{-\alpha-1}}{e^u - e^{-s(t)}}\,\mathrm{d}u
\end{equation}
provided $\alpha<0$. Compare this with the integral representation of the polylogarithmic function
\begin{equation}
		\operatorname{Li}_{\nu}(w)=\frac{w}{\Gamma(\nu)} \int_0^{\infty} \frac{t^{\nu-1}}{e^t-w}\,\mathrm{d}t, \;\; \Re(\nu)>0 .
\end{equation}
Going back to the original variable, we obtain the desired action
\begin{equation}\label{pet}
		(-h(t)D_t)^{\alpha}K_L(t) = \operatorname{Li}_{-\alpha}\left(e^{-\Phi(t)}\right) ,
\end{equation}
for $\alpha<0$. This result can be analytically continued to find that the result of the differential form is recovered.
	
We now show that this result extends to all complex values of $\alpha$. We do so by repeated integration by parts. With $s=s(t)$ we rewrite equation \eqref{pek} amenable to integration by parts,
\begin{equation}
	L^{\alpha} K_L(t)=\frac{1}{\Gamma(-\alpha)} \int_0^{\infty} \frac{u^{-\alpha-1}}{e^{u+s}-1}\,\mathrm{d}u
	\end{equation}
which is valid only for $\Re(\alpha)<0$. This restriction is brought about by the factor $u^{-\alpha-1}$ in the integrand. This restriction is mitigated by performing integration by parts with the algebraic powers of $u$ as the integrated factor each time an integration by parts is executed. With the equality $\partial_u (e^{u+s}-1)^{-1}=\partial_s (e^{u+s}-1)^{-1}$, we obtain after $n$-successive integration by parts
\begin{equation}
		L^{\alpha}K_L(t)=(-1)^n \frac{\mathrm{d}^n}{\mathrm{d}s^n} \frac{1}{\Gamma(n-\alpha)}\int_0^{\infty}\frac{u^{n-\alpha-1}}{e^{u+s}-1}\,\mathrm{d}u, \end{equation}
which is valid for $\Re(\alpha)<(n-1)$. We recognize that the integral is just the polylogarithm, so that
\begin{equation}
		L^{\alpha}K_L(t)=(-1)^n \frac{\mathrm{d}^n}{\mathrm{d}s^n} \operatorname{Li}_{n-\alpha}\left(e^{-s}\right)
\end{equation}
under the same restriction on $\alpha$. This can be evaluated using the known property of the polylogarithm function, $\mathrm{d}\operatorname{Li}_{\nu}(w)/\mathrm{d}w=\operatorname{Li}_{\nu-1}(w)/w$. From this we have $D_s \operatorname{Li}_{\nu}(e^{-s})=-\operatorname{Li}_{\nu-1}(e^{-s})$. Application of this and substituting back $s=\Phi(t)$ finally gives
\begin{equation}
	L^{\alpha}K_L(t)=\operatorname{Li}_{-\alpha}\left(e^{-\Phi(t)}\right).
\end{equation}
Clearly this result can be arbitrarily extended as far as one may desire to the right of the origin. Thus the actions of differential and integral forms are equivalent.

\subsection{Construction of the fractional trace identity}
	
The generalized spectral function corresponding to a given regularization of the trace identities is obtained by extending the contour integral representation of the trace identities to non-integral values of $m$. This requires careful consideration of the analytic properties of the function $\operatorname{Li}_{-\alpha}(e^{-\Phi(z)})$ in the complex plane. Its properties are inherited from the properties of the polylogarithmic function $\operatorname{Li}_{\nu}(w)$. For fixed $\nu$, $\operatorname{Li}_{\nu}(w)$ is analytic in the cut plane $\mathbb{C}\setminus[1,\infty)$, i.e. it has a branch point at $w=1$ and $w=\tilde{\infty}$. On the branch cut, it is continuous from below,
	\begin{equation}
		\operatorname{Li}_{\nu}(x-i 0^+)=\operatorname{Li}_{\nu}(x),\;\; x>1
	\end{equation}
 and discontinuous from above
	\begin{equation}
		\operatorname{Li}_{\nu}(x+i 0^+)=\operatorname{Li}_{\nu}
		(x) + \frac{2\pi i}{\Gamma(\nu)} \ln^{\nu-1}(x),\;\; x>1 .
	\end{equation}
Much of the analytical properties of $\operatorname{Li}_{-\alpha}(e^{-\Phi(z)})$ depend on the interaction between the zeros and singularities of $\Phi(z)$ and the singularities of $\operatorname{Li}_{\nu}(w)$ itself. 
	
We recall that we have restricted $\Phi(z)$ to being entire so that its singularity lies only at complex infinity. Now the zeros of $\Phi(z)$ are mapped into the branch point $w=1$ of $\operatorname{Li}_{\nu}(w)$. By construction $z=0$ is a zero of $\Phi(z)$ so that $z=0$ is a branch point of $\operatorname{Li}_{-\alpha}(e^{-\Phi(z)})$, which we will refer to as the principal branch point. Other branch points, the secondary ones, occur at solutions to $e^{-\Phi(z)}=1$, which are the solutions to $-\Phi(z)=2\pi k i$ for $k=\pm 1,\pm 2, \pm 3, \dots$. For entire $\Phi(z)$, these are the only singularities of $\operatorname{Li}_{-\alpha}(e^{-\Phi(z)})$ in the finite complex plane. The branch cuts must be known in order to be able to determine the proper deformation of the contour integration. Since $z=0$ is always a branch point, the contour must be modified to straddle the branch cut through the origin without intersecting any of the other branch cuts. $\operatorname{Li}_{-\alpha}(e^{-\Phi(z)})$ has no poles so that only branch cuts are to be avoided. 
Once the appropriate contour $C$ has been determined, the regulator is now given by
\begin{equation}\label{regulator}
     \mathcal{R}_L(\alpha)=\frac{1}{2\pi i} \int_C \operatorname{Li}_{-\alpha}\left(e^{-\Phi(z)}\right)\,\frac{\mathrm{d}z}{z} .
 \end{equation}
 When $\alpha=m$, a non-negative integer, the branch point at $z=0$ turns into a pole, and the integral reduces to a closed integral around the origin. Thus the regulator reduces to the regulator for the trace integral identities. 

In a while, we will show how the Riemann zeta function regularization emerges from regulator \eqref{regulator} where the regulator is $\mathcal{R}_L(\alpha)=\zeta(-\alpha)$. We can turn this expression around and express the zeta function in terms of the regulator, i.e. $\zeta(\alpha)=\mathcal{R}_L(-\alpha)$. Thus the regulator \eqref{regulator} defines a corresponding generalized zeta function given by $\mathcal{Z}_L(\alpha)=\mathcal{R}_L(-\alpha)$, and assumes the contour integral representation
\begin{equation}
		\mathcal{Z}_L(\alpha)=\frac{1}{2\pi i} \int_C \operatorname{Li}_{\alpha}\left(e^{-\Phi(z)}\right)\,\frac{\mathrm{d}z}{z},
\end{equation}
which can be uniquely extended into the entire complex plane by analytic continuation. From this we can obtain the regularized product of the positive integers,
\begin{equation}
    \prod_{n=1}^{\infty} n = \exp(-\mathcal{Z}_L'(0)).
\end{equation}

\subsection{Finite-part and the regularized limit}\label{fpvsrl}
We pointed out in Section-\ref{gentrace} that the finite-part and the regularized limit are not equivalent when continuity is imposed on the regulator. We are now in a position to see this explicitly. Recall that for integral trace identities, we have
\begin{equation}
    \left(-h(t) D_t\right)^m K_L(t)=\operatorname{Li}_{-m}\left(e^{-\Phi(t)}\right) .
\end{equation}
We regularize this by taking the finite-part as $t\rightarrow 0$. For small $t=\epsilon$,
\begin{equation}\label{pwek}
    \operatorname{Li}_{-m}\left(e^{-\Phi(\epsilon)}\right)= m! \Phi(\epsilon)^{-m-1} + \zeta(-m)+\sum_{k=1}^{\infty}(-1)^k \frac{\zeta(-m-k)}{k!} \Phi(\epsilon)^k .
\end{equation}
The finite-part is obtained by separating the terms that converge, $C_{\epsilon}$, and diverge, $D_{\epsilon}$ as $\epsilon\rightarrow 0$. The finite-part is just the limit $\lim_{\epsilon\rightarrow 0}C_{\epsilon}$. The second and the third term belong to the convergent part, but the third term vanishes in the limit because $\Phi(0)=0$. Now by hypothesis, $t=0$ is a simple zero of $\Phi(t)$ so that $\Phi(t)^{-m-1}$ has a pole of order $(m+1)$ at the origin. The regular part of the Laurent series expansion of $\Phi(t)^{-m-1}$ about the origin belongs to the convergent part $C_{\epsilon}$. It may happen that $\Phi(t)^{-m-1}$ has a constant term which contributes to the desired finite-part. Then
\begin{equation}\label{fpintegral}
    \operatorname{fp} \left[\operatorname{Li}_{-m}\left(e^{-\Phi(\epsilon)}\right)\right]_{\epsilon=0} = \zeta(-m)+m! [\epsilon^0]\!\left[\Phi(\epsilon)^{-m-1}\right] .
\end{equation}
This is precisely the regulator and it is equal to the value obtained from the contour integral representation of the regularized limit.

Now we consider the fractional case. We have for non-integer values of $\alpha$ the result
\begin{equation}
        \left(-h(t) D_t\right)^{\alpha} K_L(t)=\operatorname{Li}_{-\alpha}\left(e^{-\Phi(t)}\right) .
\end{equation}
To extract the finite part we again decompose it into its converging and diverging parts as $t$ approaches zero. Using the same expansion \eqref{pwek} with $m$ replaced by $\alpha$,
\begin{equation}\label{pwek2}
    \operatorname{Li}_{-\alpha}\left(e^{-\Phi(\epsilon)}\right)= \Gamma(1+\alpha) \Phi(\epsilon)^{-\alpha-1} + \zeta(-\alpha)+\sum_{k=1}^{\infty}(-1)^k \frac{\zeta(-\alpha-k)}{k!} \Phi(\epsilon)^k .
\end{equation}
Since $\alpha$ is a non-integer, $\Phi(\epsilon)^{-\alpha-1}$ has no constant term when expanded about $\epsilon=0$. Thus the finite-part only comes from the second term only,
\begin{equation}\label{fpfractional}
    \operatorname{fp}\left[\operatorname{Li}_{-\alpha}\left(e^{-\Phi(\epsilon)}\right)\right]_{\epsilon=0} = \zeta(-\alpha) .
\end{equation}
If we assign this as the value of the fractional trace identity, then we recover the Riemann zeta function regularization. However, the finite part \eqref{fpfractional} fails to reproduce the second term in the finite part \eqref{fpintegral}. Thus the regulator in the fractional case does not continuously reduce to the regulator for the integer case. The use of the contour integral representation of the regularized limit solves this problem.

Of course, one may choose the fractional case as the reference regularization and impose global continuity of the scheme, thereby eliminating the additional terms appearing in the integral trace identities. In the present work, however, we adopt the integral trace identities themselves as the fundamental reference regularization. This choice bears the consequence of necessarily modifying the Riemann zeta regularization.

\section{Hankel Type Regulators}\label{hankelzeta}
We now consider a class of regularizations characterized by $\Phi(z)$ satisfying the property that its reduction along the real axis, $\Phi(x)$, is monotonically increasing with positive $x$ and that $-\Phi(x)$ is positive and monotonically increasing as $x$ gets more negative; moreover, $1/\Phi(x)\rightarrow 0$ as $x\rightarrow-\infty$. 
Under these conditions, the function $\operatorname{Li}_{\alpha}(e^{-\Phi(z)})$ has a branch cut along $(-\infty,0]$, among potentially infinitely many branch cuts. Assuming that no other branch cuts cross the primary branch cut $(-\infty,0]$, the regulator is given by
\begin{equation}\label{hankel}
\mathcal{R}_L(\alpha)=\frac{1}{2\pi i} \int_{-\infty}^{(0+)} \operatorname{Li}_{-\alpha}\left(e^{-\Phi(z)}\right)\,\frac{\mathrm{d}z}{z}
\end{equation}
where the contour is the Hankel contour which starts at $-\infty$, encircles the origin once in the counterclockwise direction and returns to $-\infty$ without crossing any of the other branch cuts of $\operatorname{Li}_{\alpha}(e^{-\Phi(z)})$. The vanishing of $1/\Phi(x)$ at $-\infty$ guarantees the integrability of the integral there. We will refer to such regulator as Hankel type. 

Let us unravel equation \eqref{hankel}. We do so by deforming the Hankel contour into the keyhole contour with radius $\rho$ sufficiently small so that no other singularity than the principal branch point is enclosed. Performing the integrations along the component contours yields
 \begin{equation}\label{cont}
 \begin{split}
     &\int_{-\infty}^{(0+)} \operatorname{Li}_{-\alpha}\left(e^{-\Phi(z)}\right)\,\frac{\mathrm{d}z}{z}= \int_{-\infty}^{-\rho} \operatorname{Li}_{-\alpha}\left(e^{-\Phi(x-i 0^+)}\right)\,\frac{\mathrm{d}}{x}\\
     &\hspace{22mm}+ \int_{C_{\rho}} \operatorname{Li}_{-\alpha}\left(e^{-\Phi(z)}\right)\,\frac{\mathrm{d}}{z}+\frac{1}{2\pi i} \int_{-\rho}^{-\infty} \operatorname{Li}_{-\alpha}\left(e^{-\Phi(x+i 0^+)}\right)\,\frac{\mathrm{d}}{x}
     \end{split}
 \end{equation}
 where the first term comes from below the negative real axis, the second term from the circular path $C_{\rho}$, and the third from above the negative real axis. The integrals along the negative real axis combine to pick up a contribution from the discontinuity of the polylogarithm along its branch cut. 
 
 For infinitesimal $\epsilon$, $\Phi(x\pm i \epsilon)=\Phi(x)\pm i \Phi'(x) \epsilon$. For $x<0$, we have the condition that $\Phi'(x)>0$; then $\Phi(x\pm i 0^+)=\Phi(x)\pm i 0^+$. Or $\exp(-\Phi(x\pm i 0^+))=\exp(-\Phi(x)\mp i 0^+)=\exp(-\Phi(x))\mp i 0^+$. Combining the two integrals along the negative real axis and changing variables from $x$ to $-x$, the right hand side of equation \eqref{cont} becomes
 \begin{equation}\label{combined}
\int_{\rho}^{\infty}\left[\operatorname{Li}_{-\alpha}\left(e^{-\Phi(-x)}-i 0^+\right)-\operatorname{Li}_{-\alpha}\left(e^{-\Phi(-x)}+i 0^+\right)\right]\,\frac{\mathrm{d}x}{x}+ \int_{C_{\rho}} \operatorname{Li}_{-\alpha}\left(e^{-\Phi(z)}\right)\,\frac{\mathrm{d}z}{z} . 
 \end{equation}
 Recall that $-\Phi(-x)>0$ for $x>0$ so that $\exp(-\Phi(-x))>1$ for all $x>0$. Then using the discontinuity equation for the polylogarithm function, expression \eqref{combined} further simplifies to
 \begin{equation}
     -\frac{1}{\Gamma(-\alpha)}\int_{\rho}^{\infty} \frac{1}{(-\Phi(-x))^{1+\alpha}} \frac{\mathrm{d}x}{x} + \frac{1}{2\pi i} \int_{C_{\rho}} \operatorname{Li}_{-\alpha}\left(e^{-\Phi(z)}\right)\,\frac{\mathrm{d}z}{z}
 \end{equation}
 which is valid for $\Re(\alpha)>-1$, which covers the entire regularization region.

 Next, we look at the contour integral around $C_{\rho}$. We take $\rho$ to be sufficiently small so that the expansion \eqref{polylogexp} holds. For such $\rho$, we have
 \begin{equation}
     \operatorname{Li}_{-\alpha}\left(e^{-\Phi(z)}\right) = \Gamma(1+\alpha) \left(\Phi(z)\right)^{-\alpha-1} + \zeta(-\alpha) + \sum_{k=1}^{\infty} \frac{(-1)^k}{k!} \zeta(-\alpha-k) \Phi^k(z) .
 \end{equation}
 This is to be substituted back into the contour integral. Recall that $\Phi(z)$ has a simple zero at $z=0$. Thus $\Phi^k(z)$ has no constant term for all positive integer $k$. Then the infinite series does not contribute in the integral. Only the first two terms contribute and we have the contour integral
 \begin{equation}
     \frac{1}{2\pi i} \int_{C_{\rho}} \operatorname{Li}_{-\alpha}\left(e^{-\Phi(z)}\right)\,\frac{\mathrm{d}z}{z} = \frac{\Gamma(1+\alpha)}{2\pi i} \int_{C_{\rho}} \frac{1}{\Phi^{1+\alpha}(z)}\,\frac{\mathrm{d}z}{z} + \zeta(-\alpha) .
 \end{equation}
 Substituting this back, we obtain the following representation for the regulator,
\begin{equation}\label{rep}
\mathcal{R}_L(\alpha)=\zeta(-\alpha) +\frac{\Gamma(1+\alpha)}{2\pi i} \int_{C_{\rho}} \frac{1}{\Phi^{1+\alpha}(z)}\,\frac{\mathrm{d}z}{z}
-\frac{1}{\Gamma(-\alpha)}\int_{\rho}^{\infty} \frac{1}{(-\Phi(-x))^{1+\alpha}} \frac{\mathrm{d}x}{x}.
\end{equation}

It is clear from expression \eqref{rep} that a naive replacement of $m$ in equation \eqref{xo} with $\alpha$ without appropriately deforming the contour of integration completely omits the contribution of the third term in equation \eqref{rep}. However, the expression reduces to the trace identities \eqref{xo} for non-negative integer $\alpha=m$ because the third term vanishes since $1/\Gamma(-m)=0$ for positive integer $m$.  

The last two terms of \eqref{rep} can be combined into a single Hankel contour integral. To show, we isolate the second and third terms, factor out $\Gamma(1-\alpha)/2\pi i$, and use the identities $\Gamma(\alpha)\Gamma(1-\alpha)=\pi \csc(\pi \alpha)$ and $\sin(\pi\alpha)=(e^{i\pi\alpha}-e^{-i\pi\alpha})/2i$ to rewrite the integral in the third term as a sum of two integrals. This is followed by the substitution $x\rightarrow -x$. The second and the third terms now assume the form
\begin{equation}\label{xnp2}
\begin{split}
&\frac{\Gamma(1+\alpha)}{2\pi i}\left[ 
e^{-i\pi \alpha}\int^{-\rho}_{-\infty} \frac{1}{(-\Phi(x))^{1+\alpha}} \frac{\mathrm{d}x}{x}+e^{i\pi \alpha}\int_{-\rho}^{-\infty} \frac{1}{(-\Phi(x))^{1+\alpha}} \frac{\mathrm{d}x}{x}\right.\\
&\hspace{40mm}\left.+\int_{C_{\rho}} \frac{1}{\Phi^{1+\alpha}(z)}\,\frac{\mathrm{d}z}{z}\right] .
\end{split}
\end{equation}
From below the negative real axis, we have $(\Phi(x))^{1+\alpha}=-e^{-i\pi\alpha} (-\Phi(x))^{1+\alpha}$; and from above, $(\Phi(x))^{1+\alpha}=-e^{i\pi\alpha} (-\Phi(x))^{1+\alpha}$. Hence the first integral in \eqref{xnp2} can be expressed as an integral on top of the negative real axis, and the second beneath the same axis. These two integrals along the real axis, together with the integral around the circular path, combine into a single Hankel integral,
\begin{equation}
\frac{\Gamma(1+\alpha)}{2\pi i} \int_{-\infty}^{(0+)}\frac{1}{\Phi^{1+\alpha}(z)}\,\frac{\mathrm{d}z}{z} .\nonumber
\end{equation}
Thus under the conditions imposed on $\Phi(z)$, we have established the following result for Hankel type regulators,
\begin{equation}
\mathcal{R}_L(\alpha)=\zeta(-\alpha)+\frac{\Gamma(1+\alpha)}{2\pi i} \int_{-\infty}^{(0+)} \frac{1}{\Phi^{1+\alpha}(z)}\,\frac{\mathrm{d}z}{z}, \Re(\alpha)>-1 .
\end{equation}

Aside from the last two terms reducing to a Hankel contour integral, they can also expressed as the finite-part of a divergent integral. The third term in equation \eqref{rep} diverges as $\rho\rightarrow 0$ and it can be decomposed into convergent and divergent parts. The second term likewise diverges in the same limit and it can also be decomposed into converging and diverging parts. The diverging terms exactly cancel out, leaving the finite-part of the divergent in the third term in the limit $\rho\rightarrow 0$. This yields the finite-part integral representation of the regulator
\begin{equation}\label{fpirep}
\mathcal{R}_L(\alpha)=\zeta(-\alpha) 
-\frac{1}{\Gamma(-\alpha)}\bbint{0}{\infty} \frac{1}{(-\Phi(-x))^{1+\alpha}} \frac{\mathrm{d}x}{x}, \;\; \Re\alpha>-1 .
\end{equation}
It seems that the second term vanishes as $\alpha\rightarrow m$ for non-negative integral because $1/\Gamma(-m)=0$, leading apparently to an inconsistency with equation \eqref{rep}. However, there is no inconsistency. 

This can be seen explicitly as follows. From $\Phi(z)=\int_0^z \mathrm{d}t/h(t)$ and the condition that $1/h(t)$ is analytic at the origin, which implies $h(0)\neq 0$, we have $\Phi(z)=z/h(0) + O(z^2)$. Then $\Phi(z)$ has a simple zero at $z=0$; this allows us to write $\Phi(z)=z \phi(z)$, where $\phi(0)\neq 0$. Then $(-\Phi(-x))^{\alpha+1}=x^{\alpha+1} \phi(-x)^{\alpha+1}$, which is real for all non-negative $x$. The finite-part integral becomes
\begin{equation}
    \bbint{0}{\infty} \frac{1}{(-\Phi(-x))^{\alpha+1}}\frac{\mathrm{d}x}{x} = \bbint{0}{\infty} \frac{1}{x^{\alpha+2} \phi(-x)^{\alpha+1}}\,\mathrm{d}x .
\end{equation}
Under the conditions imposed on $h(t)$, the function $\phi(-x)^{-\alpha-1}$ is analytic in the half-line $[0,\infty)$. Then in the language of \cite{galapon2023AA}, the divergent integral is Mellin type and its finite-part can be obtained from the analytic continuation of the Mellin transform
\begin{equation}
   \mathcal{M}[\phi(-x)^{-\alpha-1},s]= \int_0^{\infty} x^{s-1}\frac{1}{\phi(-x)^{\alpha+1}}\,\mathrm{d}x ,
\end{equation}
which has a non-trivial strip of analyticity owing to the conditions imposed on $h(t)$. Let $\mathcal{M}^*(s)$ denote the meromorphic continuation of the Mellin transform to the $s$-complex plane. 

Under the condition that $\phi(-x)^{-\alpha-1}$ is analytic in the non-negative half-line, $\mathcal{M}^*(s)$ is analytic everywhere in the half-plane $\Re(s)\geq 0$ except at the integers $s=1, 2, 3, \dots$ where it either has a simple pole or a removable singularity there. For values of $s$ outside the strip of analyticity of the Mellin transform, the finite-part of the corresponding divergent Mellin-integral coincides with the value $\mathcal{M}^*(s)$ when $s$ is a regular point of $\mathcal{M}^*$.  Thus, for $s=-\alpha-1$ for non-integer $\alpha$, we have the finite-part integral
\begin{equation}
    \bbint{0}{\infty} \frac{1}{x^{\alpha+2} \phi(-x)^{\alpha+1}}\,\mathrm{d}x = \mathcal{M}^*(-\alpha-1) .
\end{equation}
When $s=-m-1$ for non-negative integer $m$, $s$ is either a removable singularity or a simple pole of $\mathcal{M}^*(s)$. When it happens that $-m-1$ is a removable singularity, $\mathcal{M}^*(-m-1)=0$ and the regularized value reduces to the zeta function regularized value. On the other hand, when $-m-1$ is a pole, there is possibly a non-zero correction to the zeta function regularization which is given by
\begin{equation}
    -\lim_{\alpha\rightarrow m}\frac{\mathcal{M}^*(-\alpha-1)}{\Gamma(-\alpha)} ,
\end{equation}
where the limit is understood in the Cauchy sense. This is exactly the value of the second term in equation \eqref{rep} in the limit $\alpha\rightarrow m$.

\begin{figure}[t]
		\label{zetareg}
		\includegraphics[scale=0.6]{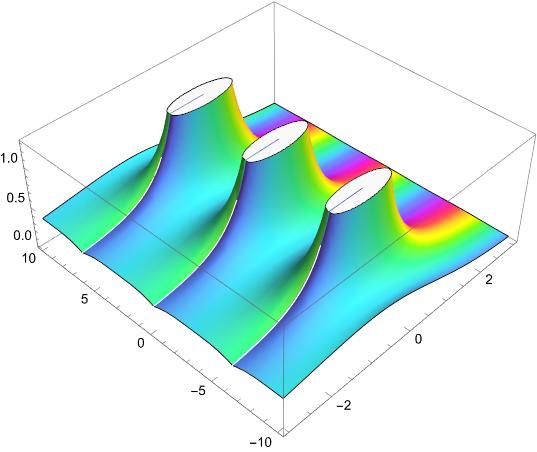}
		\caption{The plot of $\operatorname{Li}_{\alpha}(\exp(-z))$. There is one main branch cut emanating from the zeros of $z$ at $z=0$. The other branch cuts emanate from the solutions to $\exp(-z)=1$ which are not zeros of $z$.}
		\centering
\end{figure}

\subsection{Examples}
\subsubsection{The Riemann Zeta Regularization} The Riemann zeta function regulation falls under the class of Hankel type zeta functions. It corresponds to $h(t)=1$ with $\Phi(t)=t$. Indeed $\Phi(t)$ is monotonically increasing with increasing $t$ and $-\Phi(t)=-t$ is positive for $t<0$. Here $\Phi(z)=z$ is entire which has the sole zero $z=0$. The branch points of $\operatorname{Li}_{-\alpha}(e^{-z})$ are the solutions to $-z=2\pi i k$ for $k=0,\pm1,\pm2,\dots$. See figure-\ref{zetareg} for the complex plot of $\operatorname{L}_{\alpha}(e^{-z})$. The case $k=0$ corresponds to the zero of $\Phi(z)$, which is the principal branch point with the principal branch cut $(-\infty,0]$. Evaluating the first term of equation \eqref{rep} yields
\begin{equation}
\begin{split}
		\frac{\Gamma(1+\alpha)}{2\pi i} \int_{C_{\rho}} \frac{1}{\Phi^{1+\alpha}(z)}\,\frac{\mathrm{d}z}{z}=\frac{\Gamma(1+\alpha)}{2\pi i \rho^{1+\alpha}} \int_{-\pi}^{\pi} e^{-i(\alpha+1)\theta} i \mathrm{d}\theta
  = 	\frac{1}{\rho^{1+\alpha}(1+\alpha)\Gamma(-\alpha)},
  \end{split}
\end{equation}
upon using the identity $\Gamma(z)\Gamma(1-z)=\pi\csc(\pi z)$. On the other hand, the second term is given by
\begin{equation}
\begin{split}
    \frac{1}{\Gamma(-\alpha)}\int_{\rho}^{\infty} \frac{1}{(-\Phi(-x))^{1+\alpha}}\,\frac{\mathrm{d}x}{x} &= \frac{1}{\Gamma(-\alpha)} \int_{\rho}^{\infty} \frac{1}{x^{1+\alpha}}\frac{\mathrm{d}x}{x}
    =\frac{1}{\Gamma(-\alpha)} \frac{1}{\rho^{1+\alpha}(1+\alpha)}
    \end{split}
\end{equation}
Clearly the two terms cancel leaving the result 
\begin{equation}
    \mathcal{R}_L(\alpha)=\zeta(-\alpha),\;\;\operatorname{Re}(\alpha)>-1 .
\end{equation}
Extending this in the entire complex plane, we recover the well known Riemann zeta regularization. Observe that the region of validity of the equality covers all the divergent region of the sum $\sum_{n=1}^{\infty} n^{-\alpha}$, with the convergent region $\Re(\alpha)>1$. 

Incidentally, our result yields the following apparently new contour integral representation of the Riemann zeta function
\begin{equation}
	\zeta(\alpha)=\frac{1}{2\pi i} \int_{-\infty}^{0^+} \operatorname{Li}_{\alpha}\left(e^{-z}\right)\frac{\mathrm{d}z}{z} ,
	\end{equation}
valid for $\operatorname{Re}(\alpha)<1$, where the contour starts from $-\infty$ from below of the negative real axis, goes around once the origin and then back again to $-\infty$, without crossing any other branch cuts of $\operatorname{Li}_{\alpha}(e^{-z})$.
	
\subsubsection{Polynomial Regularization} 
\begin{figure}[t]
\label{tear2}
\includegraphics[scale=0.6]{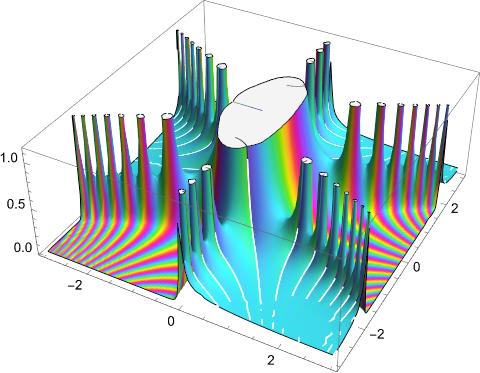}
	\caption{The plot of $\operatorname{Li}_{\alpha}(\exp(-z-z^3))$. There are three main branch cuts emanating from the zeros of $(z+z^3)$ at $z=0,\pm i$. The other branch cuts emanate from the solutions to $\exp(-z-z^3)=1$ which are not zeros of $(z+z^3)$.}
\centering
\end{figure}
Another example is the case of odd $\Phi(t)$ in $t$, i.e. $\Phi(-t)=-\Phi(t)$ where $\Phi(t)$ is monotonically increasing with $t$. This comprises a huge class of $h(t)$'s. One such is the case of $1/h(x)=(1+3x^2)$ which corresponds to $\Phi(z)=z+z^3$. The zeros of $\Phi(z)$ are $z=0,\pm i$, which are branch points of $\operatorname{Li}_{-\alpha}(e^{-\Phi(z)})$; the solution $z=0$ is the principal branch point. The other branch points are solutions to $z+z^3=-2\pi k i$ with $k=\pm 1, \pm 2, \pm 3,\dots$. These branch points are shown in figure-2. The principal branch points belong to one-fat peak, while the secondary branch points are the individual peaks.  The function $e^{-z-z^3}$ becomes unbounded in the sectors $\frac{\pi}{6}<\theta<\frac{\pi}{2}$, $\frac{5\pi}{6}<\theta<\frac{7\pi}{6}$, $\frac{3\pi}{2}<\theta<\frac{11\pi}{6}$ as $|z|\rightarrow\infty$. Along the rays $(e^{i\pi/3},e^{i\pi},e^{i5\pi/3})$ $e^{-z-z^3}$ take zero phase and approaches $\infty$ there. Then $(\infty e^{i\pi/3},\infty e^{i\pi},\infty e^{i5\pi/3})$ are branch points. The branch points along the sectors connect to these branch points asymptotically.

Clearly the contour of integration is the Hankel contour and hence we have a Hankel type generator. Then, by equation \eqref{fpirep}, the corresponding regulator is then given by
\begin{equation}\label{regreg}
   \mathcal{R}_L(\alpha)=\zeta(-\alpha)  - \frac{1}{\Gamma(-\alpha)}\bbint{0}{\infty} \frac{1}{x^{\alpha+2}(1+x^2)^{1+\alpha}} \mathrm{d}x.
\end{equation}
Using the method of Mellin transform \cite{galapon2023AA}, the finite-part integral can be evaluated from the Mellin transform
\begin{equation}
    \mathcal{M}[(1+x^2)^{-\alpha-1}; s]=\int_0^{\infty} \frac{x^{s-1}}{(1+x^2)^{\alpha +1}}\,\mathrm{d}x .
\end{equation}
The integral converges in the strip $\Re(2\alpha-s)>-2$ and $\Re(s)>0$, and is given by
\begin{equation}\label{bla}
    \mathcal{M}[(1+x^2)^{-\alpha-1}; s]=\frac{\Gamma\left(1+\alpha-\frac{s}{2}\right)\Gamma\left(\frac{s}{2}\right)}{2\Gamma\left(\alpha+1\right)} .
\end{equation}
The right hand side of equation \eqref{bla} extends the Mellin transform in the entire $s$-complex plane for a given fixed $\alpha$, which is the desired analytic continuation, $\mathcal{M}^*(s)$ of the Mellin transform. For fixed $\alpha$ in the regularization regime $\Re(\alpha)>-1$, $\mathcal{M}^*(s)$ is analytic everywhere in the half-plane $\Re(s)<0$ except at the points $s=0, -2, -4, \dots$ where it has simple poles. Then for non-integer $\alpha$ with $\Re(\alpha)>-1$, the point $s=-\alpha-1$ is a regular point of $\mathcal{M}^*(s)$. Then the desired finite-part integral in equation \eqref{regreg} is equal to $\mathcal{M}^*(s=-\alpha-1)$,
\begin{equation}
    \bbint{0}{\infty}\frac{1}{x^{\alpha+2}(1+x^2)^{\alpha+1}}\,\mathrm{d}x=\frac{\Gamma\left(-\frac{\alpha+1}{2}\right) \Gamma\left(3\frac{(\alpha+1)}{2}\right)}{2 \Gamma(\alpha+1)}
\end{equation}
Substituting this back into equation \eqref{regreg} and simplifying, we obtain the regulator
\begin{equation}\label{exreg}
   \mathcal{R}_L(\alpha)=\zeta(-\alpha)  - \frac{\Gamma(3(1+\alpha)/2) \sin(\pi\alpha/2)}{\Gamma((3+\alpha)/2)} .
\end{equation}

In the regularization regime $\Re(\alpha)>-1$, the regulator is analytic everywhere. We can now check if the condition for continuity is satisfied as $\alpha$ approaches a positive integer. Equation \eqref{exreg} reduces to
\begin{equation}
    \mathcal{R}_L(2k)=\zeta(-2k),
\end{equation}
\begin{equation}
    \mathcal{R}_L(2k-1)=\zeta(-2k+1)+(-1)^k \frac{(3k-1)!}{k!} ,
\end{equation}
in the limit for $k=1, 2, 3, \dots$. Indeed, one can confirm that application of equation \eqref{intheorem} leads to these two expressions with $\Phi(z)=z+z^3$. One can also generate these results by repeated action of the generator $L=-(1+3t^2)^{-1} D_t$ on the generalized spectral function
\begin{equation}
    K_L(t)=\frac{1}{e^{t+t^3}-1},
\end{equation}
followed by expanding the result about $t=0$ and assigning the value of the constant term in the expansion as the regularized value. 

From the regulator, we can construct the corresponding generalized zeta function via the replacement $\alpha\rightarrow -\alpha$. The result is
\begin{equation}
\mathcal{Z}_L(\alpha)=\zeta(\alpha)+\frac{\Gamma(3(1-\alpha)/2) \sin(\pi\alpha/2)}{\Gamma((3-\alpha)/2)} .
\end{equation}
And with this we get the regularized product
\begin{equation}
    \prod_{n=1}^{\infty} n = \sqrt{2\pi} \, e^{-\pi/2} ,
\end{equation}
coming from the regularization $e^{-\mathcal{Z}_L'(0)}$. For the Riemann zeta function regularization we have the regularized value $\sqrt{2\pi}$ for the same divergent product.
	
\section{Conclusion}\label{conclusion}
We have developed a new regularization of the divergent series $\sum_{n=1}^{\infty}n^{\alpha}$ for  $\mathrm{Re}\,\alpha>-1$ that yields the Reimann zeta function regularization as a special case. It can be extended to other divergent series of the form $\sum_{n=1}^{\infty} \lambda_n^{\alpha}$, where $\lambda_n$ monotonically increases with $n$. The same set of prescriptions can be applied to regulate the series: We pick a generator, $L=L(D_{t})$ such that the eigenvalue problem $L g_n(t)= \lambda_n g_n(t)$ can be meaningfully solved for all $n$; construct the generalized spectral function, $K_L(t)=\sum_{n=1}^{\infty} g_n(t)$, which defines what meaningful solution to the eigenvalue problem is; from this obtain the trace identities $\sum_{n=1}^{\infty} \lambda_n^{m}$ by $m$-repeated application of $L$ on $K_L(t)$, followed by taking the regularized limit as $t\rightarrow 0$, implemented as a contour integral; and finally, extend the integral trace identities to non-integer powers by fractional calculus under the condition that $L^{\alpha}g_n(t)=\lambda_n^{\alpha} g_n(t)$ for non-integer $\alpha$ and that the regularized fractional trace identity $\sum_{n=1}^{\infty}\lambda_n^{\alpha}$ continuously reduces to the regularized integral trace identity $\sum_{n=1}^{\infty}\lambda_n^{m}$ 
as $\alpha\rightarrow m$.   

While the generalization appears straightforward, it is in fact nuanced. Here we have demanded that the generalized spectral function has a holomorphic extension in the entire complex plane to execute a contour integral implementation of the extraction of the constant term. But the GSF $\sum_{n=1}^{\infty} e^{-\lambda_n t}$ does not necessarily have such a holomorphic extension. For example, for the case $\lambda_n=n^M$ for every positive integer $M$, the GSF $\sum_{n=1}^{\infty} e^{-n^M t}$ has the imaginary axis as a natural boundary, by the well-known Fabry gap theorem \cite{montgomery}, so that it cannot be analytically continued into the left-hand plane. This obstruction prevents the use of contour-integral methods for extracting the constant term, since such procedures require integration around the origin and hence analytic access to regions lying to the left of the imaginary axis. We defer a detailed treatment of this issue to a separate work where the finite-part integral will play a central role.

It should be noted that our treatment here is formal. This is intentional to allow the underlying ideas to emerge without the immediate constraints of full rigor. Nevertheless, the formulation can be placed on a firm analytical footing in a Banach space $\mathcal{B}_L$, where the generator $L$ is a closed sectorial operator. In this setting, the fractional powers of $L$ can be rigorously defined by means of the Balakrishnan fractional power of a sectorial operator \cite{balakrishnan}. Moreover, the generator $L$, with the spectrum $\{1,2,3,\dots\}$ in $\mathcal{B}_L$, may be regarded as an isospectral transformation (in the sense of similarity or intertwining) of the reference generator $L_0=-\mathrm{d}/\mathrm{d}t$ with the same spectrum $\{1,2,3,\dots\}$ in the Banach space $\mathcal{B}_{L_0}$. In an analogy of the classical question, “Can one hear the shape of a drum?” \cite{kak}, one may similarly ask whether the regularization can be “heard” from the spectrum alone. Our answer here is in the negative. A more detailed account of the Banach space formulation will be given in a separate work.

The scheme constitutes a broad class of regularizations, opening the possibility of resolving potential inconsistencies in the use of the standard Riemann zeta regularization. For example, in the case of the fermion gas in a box, we have seen that Riemann zeta regularization predicts the absence of restoring force owing from the fact that $\zeta(-2)=0$ which does not agree with the intuitive expectation of the presence of a restoring force. Our regularization may provide a desirable non-vanishing value, which from our results, may assume the value
\begin{equation}
    \sum_{n=1}^{\infty}n^2 = \frac{1}{4} \left(h^{(3)}(0) h(0)^2+2 h(0) h'(0) h''(0)\right)
\end{equation}
where $h(t)$ determines the regularization. Clearly, there are infinitely many possible regularization of $\sum_{n=1}^{\infty} n^2$ corresponding to the infinitely many possible choices for $h(t)$. It may have the value zero, for example, with the choice $h(t)=(1+3t^2)^{-1}$; a negative value, for example, with the choice $h(t)=(1+2t)^{-1}$ that yields the regularization $\sum_{n=1}^{\infty}n^2=-20$; a positive value, for example, with the choice $h(t)=(1+2 t +3 t^2)^{-1}$ that yields the regularization $\sum_{n=1}^{\infty}n^2 = 4$.  The first case predicts the absence of any force after the displacement of the partition; the second, presence of a repulsive force; the third, the presence of a restoring force. The case for the third case is compelling but only experiment can decide which is the correct regularization to be applied. 

The prodigious number of available regularizations 
may appear to represent arbitrary mathematical freedom rather than any meaningful physical parameterization. However, physical considerations ultimately guide the selection of $L$. Taking hint from the results of Kolomeisky, Straley and Timmins on the one-dimensional Casimir effect \cite{kolomeisky}, where full energy of the fermion gas in a box with a partition decomposes into a smooth regularized envelope plus a correction imposed by the integer constraint on particle number, one may view $L$ as encoding information not visible to the spectrum. Different physical settings would correspond to different generators within the isospectral family, with the choice fixed by what the system requires beyond its spectrum, e.g. additional constraints and boundary data. Under this reading, the proliferation of schemes is not arbitrary mathematical freedom but a parametrization of physical inputs, and the standard Riemann zeta scheme corresponds to the case in which no such additional information is present beyond the spectrum. We leave a systematic development of this perspective to future work.

Finally, the existence of alternative regularizations invites speculation on a connection between regularization and the equation of motion of the system under consideration. Suppose that the system has  the Hamiltonian $H_{\vec{x}}$ with the equation of motion  $H_{\vec{x}}\psi(\vec{x},t)=L_{t} \psi(\Vec{x},t)$, and that $H_{\vec{x}}$ solves the eigenvalue problem $H_{\vec{x}}\psi_n(\Vec{x})=\lambda_n\psi_n(\Vec{x})$. Separating variables, $\psi(\vec{x},t)=\psi_n(\vec{x}) g_n(t)$, reduces the equation of motion to  $L_{t} g_n(t)=\lambda_n g_n(t)$, which is precisely the eigenvalue problem that defines a regularization corresponding to a generator $L_{t}$. The corresponding propagator $K(\Vec{x},\vec{x}',t)=\sum_{n=1}^{\infty} g_n(t) \psi_n(\Vec{x})\psi_n^*(\vec{x}')$ has the trace $K_L(t)$ from which everything follows. This suggests a direct correspondence between a choice of regularization and a specific equation of motion. A notable consequence is that the inverse problem of finding the correct regularization for a given divergent sum becomes the problem of identifying the correct equation of motion for the system that produces the sum. Thus one may, in principle, discover the right equation of motion by discovering the right regularization, and conversely, one may discover the right regularization by discovering the equation of motion.

\end{document}